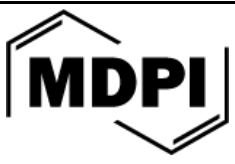

# Artificial Intelligence in Lifelong Learning: Opportunities and Challenges in Adult Education Policy

Andresa Theodora, Nikolaos Tselios

**Abstract:** Artificial intelligence (AI) is increasingly reshaping lifelong learning by introducing new possibilities for personalized, flexible, and data-informed educational practices. In the field of adult education, AI has gained particular importance as learners are expected to continuously update their knowledge and skills in response to rapid technological, economic, and social change. This paper examines the role of AI in adult education policy, with a focus on both its opportunities and its challenges. It discusses how AI can support personalized learning, intelligent tutoring, learning analytics, and workforce development, while also contributing to greater accessibility, scalability, and policy responsiveness. At the same time, the paper highlights significant concerns related to the digital divide, data privacy, algorithmic bias, over-reliance on technology, and the readiness of educators and institutions to integrate AI effectively. Drawing on contemporary literature and international policy frameworks, the paper argues that AI should not be approached simply as a technological solution, but as a socio-technical and ethical issue that requires careful governance. It concludes that the successful integration of AI into lifelong learning depends on balanced adult education policies that promote inclusion, transparency, human-centered pedagogy, and responsible innovation.

## 1 Introduction

Artificial intelligence (AI) has moved rapidly from being a specialized technological field to becoming a visible part of everyday educational practice. In education, AI generally refers to computational systems that can perform functions associated with human cognition, such as pattern recognition, prediction, language generation, feedback, and decision support. Recent international guidance presents AI not simply as a tool, but as a sociotechnical development that is reshaping how knowledge is accessed, produced, and evaluated in learning environments (OECD, 2024a; UNESCO, 2025). In adult education, this shift is especially significant because AI-based systems are increasingly used for personalized learning pathways, automated tutoring, skills assessment, and support for flexible study across formal, non-formal, and informal contexts (OECD, 2024b; Wang et al., 2024).

At the same time, the discussion of AI in education cannot be separated from the broader concept of lifelong learning. Lifelong learning refers to the continuous development of knowledge, skills, values, and competences throughout life. In contemporary societies marked by digital transformation, labour-market instability, population ageing, and growing social inequality, adult learning is no longer viewed as a secondary policy concern. It is increasingly treated as a central condition for employability, democratic participation, social inclusion, and personal fulfilment (OECD, 2024; UNESCO, 2024a). From this perspective, adult education policy must respond not only to the expansion of digital tools, but also to the changing expectations placed on citizens to reskill and adapt repeatedly across the life course.

This policy relevance is reflected in current international frameworks. The European Commission’s Digital Education Action Plan 2021-2027 promotes high-quality, inclusive, and accessible digital education, while UNESCO has emphasized that AI governance in education should be guided by human rights, inclusion, transparency, and learner protection (European Commission, 2025; UNESCO, 2025). These frameworks show that AI is not only a matter of innovation, but also a matter of regulation, ethics, and public responsibility. For adult education systems, the issue is particularly urgent because adult learners often enter learning spaces with unequal digital access, diverse educational backgrounds, and immediate employment-related pressures.

Against this background, the present paper examines why AI matters in adult education policy. Its main aim is to explore both the opportunities and the challenges associated with the integration of AI into lifelong learning. More specifically, it seeks to analyse how AI may enhance access, personalization, and policy responsiveness, while also raising concerns related to inequality, data privacy, bias, and institutional readiness. To address these issues, the paper is structured in six parts. Following this introduction, the second section outlines the conceptual and theoretical framework. The third examines key applications of AI in lifelong learning. The fourth discusses the main opportunities for adult education policy, while the fifth analyses major challenges and risks. The final section considers the broader implications for policy and future practice.

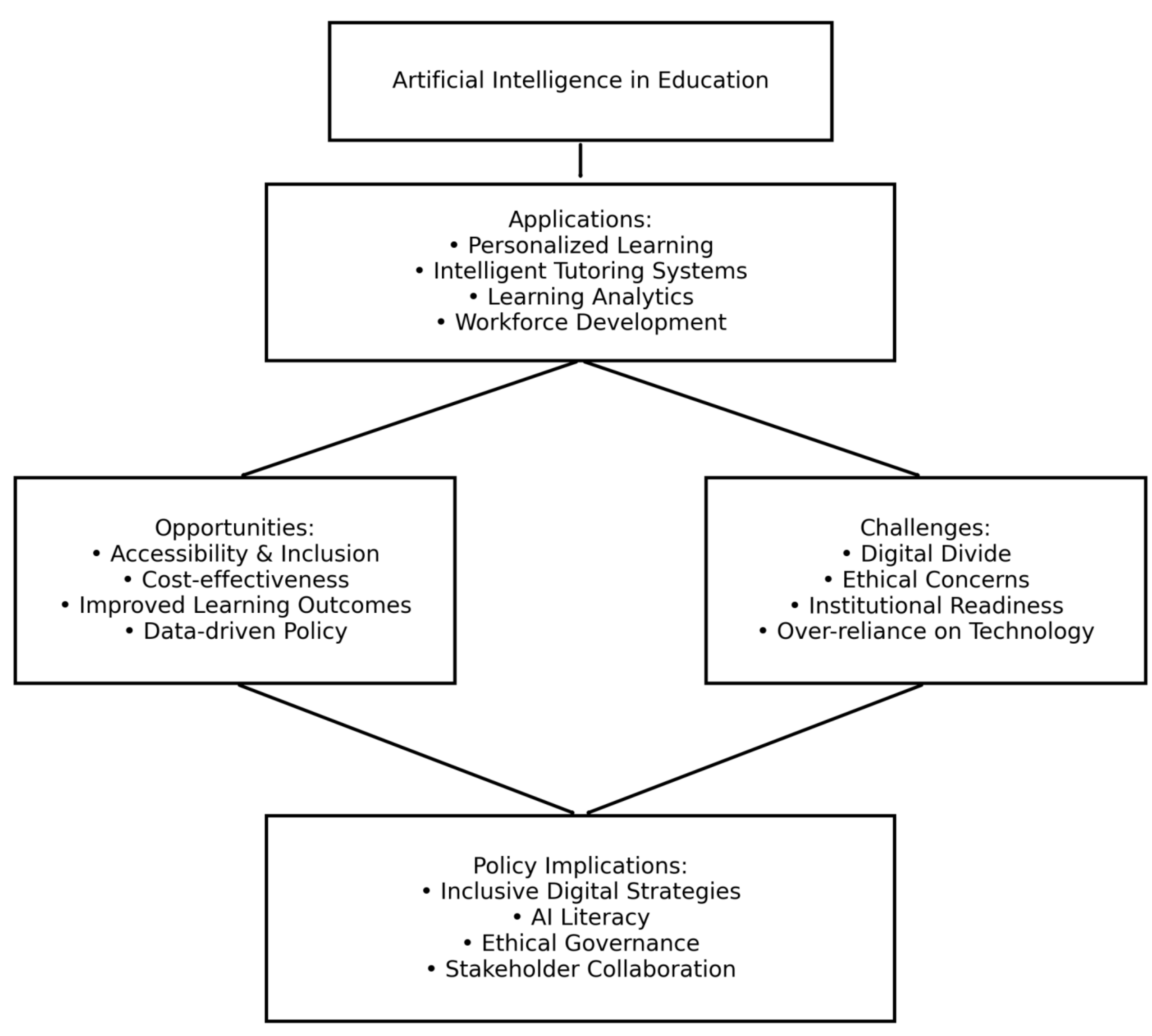

***Scheme 1.*** Conceptual framework of artificial intelligence in lifelong learning and adult education policy

## 2 Conceptual and Theoretical Framework

Understanding the role of artificial intelligence (AI) in lifelong learning requires a multidimensional framework that connects technological developments with established theories of adult learning and broader socio-educational perspectives. Rather than viewing AI as a neutral tool, recent literature emphasizes its position within complex educational ecosystems, where pedagogy, policy, and technology interact dynamically (Cukurova, 2024; OECD, 2026).

### 2.1 Artificial Intelligence as a Socio-Technical Educational System

AI in education is increasingly conceptualized as part of a broader socio-technical system, rather than as a set of isolated digital tools. This perspective highlights how AI technologies are shaped by-and in turn reshape-social practices, institutional structures, and educational values. Contemporary research suggests that AI can be understood through three main paradigms: AI-directed learning, AI-supported learning, and AI-empowered learning. These models differ in the degree of learner autonomy and the extent to which AI interacts with human cognition, ranging from automated content delivery to collaborative and learner-centered environments .

More recent theoretical approaches go further by introducing the concept of hybrid intelligence, where human and artificial intelligence complement one another. In this view, AI does not replace human thinking but extends it, supporting decision-making, reflection, and problem-solving processes. This aligns with the idea that effective educational systems should combine human judgment with data-driven insights, particularly in complex adult learning contexts (Cukurova, 2024; OECD, 2026).

### 2.2 Adult Learning Theories and AI Integration

The integration of AI into lifelong learning must also be examined through the lens of adult learning theory. Unlike traditional pedagogy, adult education is grounded in andragogy, which emphasizes self-direction, prior experience, goal orientation, and the immediate applicability of knowledge. Adult learners typically engage in education with specific personal or professional objectives, often balancing learning with work and family responsibilities.

AI technologies appear particularly compatible with these principles. Adaptive learning systems, intelligent tutoring platforms, and learning analytics can provide personalized pathways, flexible pacing, and real-time feedback-features that correspond closely with adult learners' needs for autonomy and relevance. Empirical research indicates that such systems can enhance engagement, motivation, and learning outcomes, especially when aligned with learners' goals and prior knowledge .

However, theoretical discussions also caution against overly technocentric interpretations. Adult learning is not only cognitive but also social and experiential. From a constructivist perspective, knowledge is co-constructed through interaction, reflection, and dialogue. Therefore, AI-supported learning environments must be designed in ways that preserve collaboration, critical thinking, and meaningful human interaction, rather than reducing learning to automated processes.

### 2.3 Lifelong Learning in the Digital Age

Lifelong learning itself has evolved conceptually in response to digital transformation. It is no longer limited to continuous education but is increasingly understood as a dynamic and adaptive process, closely linked to employability, innovation, and social participation. AI plays a central role in this transformation by enabling scalable, personalized, and data-driven learning opportunities across different life stages (OECD, 2024c; Wang et al., 2024).

At the same time, the expansion of AI in lifelong learning raises important theoretical questions about agency, autonomy, and cognition. Emerging research highlights potential risks such as cognitive offloading, reduced learner agency, and dependence on algorithmic systems. These concerns suggest that AI should not only support efficiency but also foster critical engagement and independent thinking (Favero et al., 2026).

### 2.4 AI Literacy and Educational Policy Frameworks

A key concept linking theory and policy is AI literacy, which refers to the knowledge, skills, and attitudes required to understand, use, and critically evaluate AI systems. Recent frameworks developed by international organizations emphasize that AI literacy is not purely technical; it also includes ethical awareness, critical thinking, and the ability to engage responsibly with digital technologies .

From a policy perspective, this expands the goals of adult education beyond skill acquisition to include empowerment and informed participation in AI-driven societies. It also highlights the shared responsibility of educators, institutions, and policymakers in ensuring that AI integration supports inclusive and equitable learning environments.

| Dimension | Key Components | Description |
|---|---|---|
| **Artificial Intelligence in Education** | Core Concept | Refers to the use of computational systems capable of performing cognitive functions such as prediction, personalization, and decision support in learning environments. |
| **Applications** | Personalized Learning | Adaptive systems that tailor content, pace, and pathways to individual learner needs. |
| | Intelligent Tutoring Systems | AI-driven tools providing real-time feedback, guidance, and assessment. |
| | Learning Analytics | Data-driven monitoring and evaluation of learner performance and engagement. |
| | Workforce Development | AI-supported identification of skills gaps and alignment of learning with labour market needs. |
| **Opportunities** | Accessibility & Inclusion | Expansion of learning access for diverse and underserved populations. |
| | Cost-effectiveness | Reduction of costs through automation and scalable learning solutions. |
| | Improved Learning Outcomes | Enhanced engagement, personalization, and retention. |
| | Data-driven Policy | Use of analytics to inform evidence-based educational decision-making. |
| **Challenges** | Digital Divide | Inequalities in access to technology and digital skills. |
| | Ethical Concerns | Issues related to privacy, data protection, and algorithmic bias. |

| Dimension | Key Components | Description |
|---|---|---|
| | Institutional Readiness | Need for teacher training and organizational adaptation. |
| | Over-reliance on Technology | Risk of reducing human interaction and critical engagement. |
| **Policy Implications** | Inclusive Digital Strategies | Policies ensuring equitable access to AI-supported learning. |
| | AI Literacy | Development of critical and technical competencies for engaging with AI. |
| | Ethical Governance | Implementation of transparency, accountability, and fairness in AI use. |
| | Stakeholder Collaboration | Cooperation among governments, educators, institutions, and learners. |

***Table 1*****:** Conceptual Framework of Artificial Intelligence in Lifelong Learning and Adult Education Policy

## 3 Applications of AI in Lifelong Learning

The integration of artificial intelligence (AI) into lifelong learning has moved beyond experimental use and is now embedded in a wide range of educational practices, particularly in adult education. What distinguishes AI from earlier digital tools is not simply automation, but its capacity to process large volumes of data, recognize patterns, and adapt learning experiences in real time. This section examines some of the most prominent applications of AI in lifelong learning, with particular attention to their relevance for adult learners.

### 3.1 Personalized Learning

One of the most widely discussed applications of AI in education is personalized learning. Adaptive platforms use algorithms to adjust content, pacing, and difficulty according to individual learners' needs, preferences, and prior knowledge. For adult learners-who often return to education with diverse experiences and varying skill levels-this level of customization is particularly valuable. AI systems can identify gaps in knowledge and propose targeted learning activities, thereby supporting more efficient and meaningful learning pathways (Holmes et al., 2022; Zawacki-Richter et al., 2019).

In addition, AI-driven recommendation systems function in ways similar to those used in digital media platforms, suggesting courses, resources, or learning trajectories based on user behaviour and goals. These systems can help adult learners navigate increasingly complex learning ecosystems, where the abundance of available content can otherwise become overwhelming. However, while personalization enhances relevance, it also raises questions about learner autonomy and the transparency of algorithmic decisions (OECD, 2024).

### 3.2 Intelligent Tutoring Systems

Intelligent Tutoring Systems (ITS) represent another important application of AI in lifelong learning. These systems simulate aspects of human tutoring by providing automated feedback, guidance, and assessment. Unlike traditional e-learning platforms, ITS can respond dynamically to learner input, offering explanations, hints, or corrective feedback tailored to individual performance.

For adult learners, who often engage in self-directed learning outside formal educational settings, such systems can provide continuous support without the need for constant instructor presence. This is particularly relevant in distance and online learning environments, where access to immediate feedback is often limited. Research suggests that ITS can improve learning outcomes, especially in domains that require structured knowledge acquisition, such as language learning or technical skills (Holmes et al., 2022; Woolf, 2021).

At the same time, it is important to recognize that automated feedback cannot fully replace the pedagogical and emotional dimensions of human interaction. Adult learning frequently involves reflection, discussion, and the negotiation of meaning-elements that remain difficult to replicate through AI alone.

### 3.3 Data Analytics and Learning Assessment

AI has also transformed how learning is monitored and assessed through the use of learning analytics. These systems collect and analyse data on learner behaviour, engagement, and performance, enabling educators and institutions to track progress in real time. For adult education providers, this can support more responsive and flexible forms of assessment, moving beyond traditional exams toward continuous evaluation.

A particularly significant development is the use of predictive analytics to identify learners at risk of dropping out. By analysing patterns such as reduced participation or declining performance, AI systems can generate early warnings and trigger targeted interventions. This is especially important in adult education, where dropout rates can be high due to competing responsibilities such as work and family (Siemens & Baker, 2022).

Nevertheless, the increasing reliance on data-driven assessment raises concerns about privacy, data ownership, and the potential for surveillance. These issues highlight the need for clear ethical guidelines and transparent data practices in adult education policy.

### 3.4 AI in Workforce Development

AI is also playing a central role in workforce development, particularly in the context of upskilling and reskilling. As labour markets evolve rapidly due to technological change, adult learners are increasingly required to update their skills throughout their careers. AI-powered platforms can support this process by identifying skill gaps, recommending relevant training, and aligning learning opportunities with labour market demands (World Economic Forum, 2023).

For example, AI systems can analyse job market data to identify emerging skills and suggest personalized learning pathways that correspond to these trends. This creates a closer link between education and employment, making lifelong learning more responsive to economic needs. However, this alignment also raises critical questions about the purpose of adult education, particularly whether it should primarily serve labour market demands or broader goals such as personal development and civic participation.

# 4 Opportunities of AI in Adult Education Policy

The growing use of AI in lifelong learning presents significant opportunities for adult education policy. While challenges remain, policymakers increasingly recognize the potential of AI to enhance access, improve learning outcomes, and support more effective governance of education systems.

## 4.1 Increased Accessibility and Inclusion

One of the most promising aspects of AI is its potential to expand access to learning opportunities. Digital platforms powered by AI can reach learners who are geographically isolated, economically disadvantaged, or otherwise excluded from traditional education systems. For example, AI-driven translation tools and speech recognition technologies can support learners with linguistic barriers, while adaptive interfaces can accommodate different learning needs and abilities (UNESCO, 2024).

For adult learners, flexibility is often a key requirement. AI enables learning to take place at any time and in various formats, making it easier to combine education with work and family responsibilities. From a policy perspective, this supports more inclusive and equitable education systems. However, it also requires addressing the digital divide, as unequal access to technology can reinforce existing inequalities.

## 4.2 Cost-Effectiveness and Scalability

AI technologies also offer the potential to reduce the cost of delivering education and to scale learning opportunities to larger populations. Automated systems can handle tasks such as assessment, feedback, and administrative support, allowing institutions to allocate resources more efficiently. This is particularly relevant in adult education, where funding is often limited and demand is increasing.

Scalability is another important advantage. AI-powered platforms can deliver learning experiences to large numbers of users simultaneously, without compromising personalization. This creates opportunities for governments to expand lifelong learning initiatives at a national or even global level (OECD, 2024). However, cost-effectiveness should not come at the expense of quality, and policy frameworks must ensure that technological efficiency is balanced with pedagogical integrity.

### 4.3 Enhancement of Learning Outcomes

AI has the potential to improve learning outcomes by increasing engagement, motivation, and retention. Personalized content, interactive learning environments, and real-time feedback can make learning more relevant and responsive to individual needs. For adult learners, who are often motivated by specific goals, such features can enhance persistence and success.

Real-time feedback is particularly valuable, as it allows learners to identify and address misunderstandings immediately. This can support more effective learning processes and reduce frustration, especially in self-directed learning contexts. Empirical studies suggest that AI-supported learning environments can lead to measurable improvements in performance, although outcomes depend on the quality of design and implementation (Holmes et al., 2022).

### 4.4 Policy Innovation and Data-Driven Decision Making

Finally, AI creates new possibilities for policy innovation through the use of data-driven decision making. Learning analytics can provide policymakers with detailed insights into participation patterns, learning outcomes, and system performance. This enables more evidence-based approaches to policy design, implementation, and evaluation.

For example, data collected from AI systems can help identify which groups are underrepresented in adult education, which programmes are most effective, and where resources should be allocated. This can improve both efficiency and equity in education systems. At the same time, it raises important questions about data governance, transparency, and accountability, which must be addressed through clear regulatory frameworks (OECD, 2024; UNESCO, 2025).

## 5 Challenges and Risks

While artificial intelligence (AI) offers considerable promise for lifelong learning, its integration into adult education systems is not without significant challenges. These challenges are not merely technical; they are deeply social, ethical, and political, requiring careful consideration within policy frameworks.

### 5.1 Digital Divide and Inequality

One of the most persistent concerns is the digital divide, which continues to shape access to education in unequal ways. Although AI-powered platforms can expand learning opportunities, they also depend on reliable internet access, digital devices, and basic technological literacy. For many adult learners-particularly those in rural areas, low-income groups, or marginalized communities-such resources are not guaranteed. As a result, AI risks reinforcing existing inequalities rather than reducing them (UNESCO, 2024; van Dijk, 2020).

Beyond access to infrastructure, there is also a skills gap that affects adult learners' ability to engage meaningfully with AI-supported environments. Digital competence varies widely across populations, and many adults may lack the confidence or experience required to navigate complex digital systems. This gap is especially evident among older learners, who may face additional barriers related to prior educational experiences and limited exposure to technology (OECD, 2024). Addressing inequality in this context therefore requires more than technological provision; it demands targeted support for digital literacy and inclusive design.

### 5.2 Ethical Concerns

The increasing use of AI in education raises important ethical concerns, particularly in relation to data privacy and surveillance. AI systems often rely on the collection and analysis of large amounts of personal data, including learning behaviours, performance metrics, and sometimes even biometric information. While such data can improve personalization and system efficiency, it also creates risks related to unauthorized access, misuse, or lack of informed consent (Williamson &

Eynon, 2020).

In addition, concerns about algorithmic bias have become central to debates on AI. AI systems are trained on datasets that may reflect existing social inequalities, which can lead to biased outcomes in areas such as assessment, recommendations, or learner profiling. In adult education, this could result in unequal opportunities or reinforcement of stereotypes, particularly for vulnerable groups. These issues highlight the importance of transparency, accountability, and ethical oversight in the development and use of AI technologies (Holmes et al., 2022).

### 5.3 Teacher and Institutional Readiness

Another critical challenge relates to the preparedness of educators and institutions. The effective integration of AI into adult education requires not only technological infrastructure but also professional development for educators. Teachers need to develop new competencies, including digital literacy, data interpretation, and the ability to critically evaluate AI tools. Without adequate training, there is a risk that AI will be used superficially or ineffectively (Redecker & Punie, 2017).

Institutional readiness is equally important. Educational organizations must adapt their structures, curricula, and support systems to accommodate new forms of learning. However, resistance to change is common, particularly when technological innovations are perceived as threatening professional roles or increasing workload. Such resistance should not be dismissed as reluctance but understood as a response to uncertainty and the need for meaningful support during transitions.

### 5.4 Over-reliance on Technology

A further concern is the potential over-reliance on AI technologies. While automation can enhance efficiency, excessive dependence on digital systems may lead to a reduction in human interaction, which remains a central element of adult learning. Dialogue, collaboration, and emotional support are essential components of meaningful educational experiences, and these cannot be fully replicated by AI systems.

There is also a broader risk of dehumanizing education, where learning becomes reduced to measurable outputs and algorithmic processes. Such an approach may overlook the social, cultural, and emotional dimensions of learning, which are particularly important in adult education contexts. Scholars have cautioned that education should not be driven solely by efficiency and performance metrics but should also support critical thinking, creativity, and personal development (Biesta, 2022).

### 5.5 Policy and Regulatory Challenges

Finally, the rapid development of AI poses significant policy and regulatory challenges. In many contexts, there is a lack of clear frameworks governing the use of AI in education. This creates uncertainty regarding issues such as data protection, accountability, and quality assurance. Policymakers often struggle to keep pace with technological change, resulting in gaps between innovation and regulation (UNESCO, 2025).

Governance is particularly complex because AI systems are often developed by private companies, while education remains a public responsibility. This raises questions about control, transparency, and the role of commercial interests in shaping educational practices. Ensuring accountability requires collaboration between governments, institutions, and technology providers, as well as the development of robust regulatory mechanisms.

## 6 Implications for Adult Education Policy

Given both the opportunities and challenges associated with AI, there is a clear need for thoughtful and forward-looking policy responses. Adult education policy must balance innovation with inclusion, efficiency with ethics, and technological advancement with human-centered values.

### 6.1 Policy Recommendations

A key priority is the development of inclusive digital strategies that ensure equitable access to AI-supported learning. This includes investment in infrastructure, such as broadband connectivity and digital devices, as well as targeted initiatives to support disadvantaged groups. Equally important is the promotion of digital literacy, enabling adult learners to engage confidently and critically with AI technologies (OECD, 2024).

Investment in training and professional development is also essential. Educators must be supported in developing the skills needed to integrate AI effectively into their teaching practices. This includes not only technical skills but also pedagogical and ethical competencies.

### 6.2 Role of Stakeholders

The successful integration of AI into lifelong learning requires the active involvement of multiple stakeholders. Governments play a central role in setting policy directions, regulating technologies, and ensuring equity. Educational institutions are responsible for implementing AI in ways that align with pedagogical goals and learner needs.

Educators act as mediators between technology and learners, shaping how AI is used in practice. Finally, learners themselves should not be seen as passive recipients but as active participants, whose experiences and perspectives can inform the design and evaluation of AI systems. Collaboration among these stakeholders is essential for creating sustainable and inclusive education systems.

### 6.3 Future Policy Directions

Looking ahead, there is a growing need for ethical AI frameworks that guide the development and use of technology in education. Such frameworks should emphasize transparency, fairness, accountability, and respect for human rights. International organizations have already begun to outline these principles, but their implementation remains uneven (UNESCO, 2025).

Another important direction is the development of lifelong learning ecosystems, where formal, non-formal, and informal learning opportunities are interconnected. AI can support this by enabling flexible pathways, recognizing prior learning, and facilitating transitions between different stages of education and employment. However, achieving this vision requires coordinated policy efforts and long-term investment.

## 7 Conclusion

The integration of artificial intelligence into lifelong learning represents both a significant opportunity and a complex challenge for adult education policy. On the one hand, AI has the potential to transform learning by making it more personalized, accessible, and responsive to individual and societal needs. On the other hand, it raises critical issues related to inequality, ethics, institutional capacity, and governance.

This paper has highlighted that the impact of AI is not determined solely by technological capabilities but by the ways in which it is implemented and regulated. Opportunities such as improved access, enhanced learning outcomes, and data-driven policymaking must be carefully balanced against risks including the digital divide, algorithmic bias, and the erosion of human-centered educational values.

Ultimately, the future of AI in lifelong learning depends on the ability of policymakers, educators, and institutions to adopt a balanced and critical approach. Rather than viewing AI as a solution in itself, it should be understood as a tool that can support broader educational goals, provided that it is guided by principles of inclusion, ethics, and social responsibility. In this sense, the challenge is not only to innovate, but to ensure that innovation contributes to more equitable and meaningful learning opportunities for all.

**Author Contributions:** Conceptualization, A.K., M.M.; methodology,

A.K.; software, A.K.; validation, A.K., M.M.; formal analysis, A.K.; investigation, A.K.; resources, A.K., M.M.,; data curation, A.K.; writing-original draft preparation, A.K.; writing-review and editing, A.K.; visualization, A.K.; supervision, A.K.; project administration, A.K.; funding acquisition, M.M. All authors have read and agreed to the published version of the manuscript.

**Funding:** This research received no external funding.

**Conflicts of Interest:** The authors declare no conflicts of interest.

## References

Biesta, G. (2022). *World-centred education: A view for the present*. Taylor & Francis. https://www.scribd.com/document/618351111/Gert-Biesta-2021-World-Centred-Education-A-View-for-the-Present

Cukurova, M. (2024). The interplay of learning, analytics, and artificial intelligence in education: A vision for hybrid intelligence. *Computers and Education: Artificial Intelligence*. https://bera-journals.onlinelibrary.wiley.com/doi/10.1111/bjet.13514

European Commission. (2025). Digital Education Action Plan (2021-2027). *European Education Area*. https://education.ec.europa.eu/focus-topics/digital-education/actions

Favero, L., Pérez-Ortiz, J. A., Käser, T., & Oliver, N. (2026). AI in education beyond learning outcomes: Cognition, agency, emotion, and ethics. *arXiv*. https://arxiv.org/abs/2602.04598

Holmes, W., Bialik, M., & Fadel, C. (2019). *Artificial intelligence in education: Promises and implications for teaching and learning*. Center for Curriculum Redesign. https://curriculumredesign.org/our-work/artificial-intelligence-in-education/#1445978738683-c9203bf2-2551

OECD. (2024a). *Artificial intelligence and education and skills*. Organisation for Economic Co-operation and Development. https://www.oecd.org/en/topics/artificial-intelligence-and-education-and-skills.html

OECD. (2024b). *Readying adult learners for innovation: Reskilling and upskilling for the future*. Organisation for Economic Co-operation and Development. https://www.oecd.org/content/dam/oecd/en/publications/reports/2024/07/readying-adult-learners-for-innovation_77f4315f/85748b7b-en.pdf

OECD. (2024c). *Reimagining education: Realising potential*. Organisation for Economic Co-operation and Development. https://www.oecd.org/content/dam/oecd/en/publications/reports/2024/04/reimagining-education-realising-potential_fa0b99ab/b44e2c39-en.pdf

OECD. (2026). *Digital education outlook 2026*. Organisation for Economic Co-operation and Development. https://www.oecd.org/content/dam/oecd/en/publications/reports/2026/01/oecd-digital-education-outlook-2026_940e0dd8/062a7394-en.pdf

Redecker, C. (2017). *European framework for the digital competence of educators: DigCompEdu* (Y. Punie, Ed.). Publications Office of the European Union. https://publications.jrc.ec.europa.eu/repository/handle/JRC107466

Ritter, L. A. (2025). Adult learners' perceptions of AI-supported learning tools. *International Journal of Research and Review, 12*(12). https://www.ijrrjournal.com/IJRR_Vol.12_Issue.12_December2025/IJRR44.pdf

Siemens, G., & Baker, R. S. J. d. (2012). Learning analytics and educational data mining: Towards communication and collaboration. In *Proceedings of the 2nd International Conference on Learning Analytics and Knowledge*. 252-254. ACM. https://learninganalytics.upenn.edu/ryanbaker/Chapter12BakerSiemensv3.pdf?utm_source=chatgpt.com

UNESCO. (2024). *AI and education: Protecting the rights of learners*. UNESCO. https://www.unesco.org/en/articles/ai-and-education-protecting-rights-learners

UNESCO. (2025). *AI and education: Guidance for policy-makers*. UNESCO. https://www.unesco.org/en/articles/ai-and-education-guidance-policy-makers

van Dijk, J. (2020). *The digital divide*. Polity. https://www.academia.edu/86595666/Dijk_Jan_2020_The_digital_divide_Cambridge_UK_Polity_208_pp_17_99_paperback_ISBN_9781509534456_

Wang, S., Liu, Y., Xu, Y., Yang, H., Zhu, T., & Zeng, Q. (2024). Artificial intelligence in education: A systematic literature review. *Expert Systems with Applications, 252*, Article 124167. https://www.sciencedirect.com/science/article/pii/S0957417424010339

Williamson, B., & Eynon, R. (2020). Historical threads, missing links, and future directions in AI in education. *Learning, Media and Technology, 45*(3), 223-235. https://ora.ox.ac.uk/objects/uuid:ccba00ff-ddc0-4ffd-91ca-5670ece5414e/files/rf7623c73p

Woolf, B. P. (2009). *Building intelligent interactive tutors: Student-centered strategies for*

*revolutionizing e-learning*. Morgan Kaufmann.
https://thuvienso.dau.edu.vn:88/bitstream/DHKTDN/6557/1/Building%20Intelligent%20Interactive%20Tutors.6023.pdf

World Economic Forum. (2023). *The future of jobs report 2023*. WEF.
https://www3.weforum.org/docs/WEF_Future_of_Jobs_2023.pdf

Zawacki-Richter, O., Marín, V. I., Bond, M., & Gouverneur, F. (2019). Systematic review of research on artificial intelligence applications in higher education -where are the educators? *International Journal of Educational Technology in Higher Education, 16*, Article 39.
https://discovery.ucl.ac.uk/10176703/1/Zawacki-Richter%20et%20al%20%282019%29%20-%20Systematic%20review%20of%20research%20on%20AI%20applications%20in%20HE.pdf